\documentclass[11pt]{article}
\usepackage{epsfig}
\def\MET{\mbox{${\hbox{$E$\kern-0.6em\lower-.1ex\hbox{/}}}_T$}} 

\def\MP{\mbox{$M_P$}}
\def\mp{\mbox{$M_P$}\ }
\def\TH{\mbox{$T_H$}\ }
\def\mbh{\mbox{$M_{\rm BH}$}\ }     
\def\MBH{\mbox{$M_{\rm BH}$}}         
\def\MET{\mbox{${\hbox{$E$\kern-0.6em\lower-.1ex\hbox{/}}}_T$}} 
\def\met{\mbox{${\hbox{$E$\kern-0.6em\lower-.1ex\hbox{/}}}_T$}\ } 
\def\ifb{fb$^{-1}$}                     

\title{\Large\bf Black Holes at Future Colliders and Beyond: a Review}

\author{Greg Landsberg\\
\normalsize
Brown University, Department of Physics, 182 Hope St, Providence, RI 02912, USA\\
\normalsize
E-mail: landsberg@hep.brown.edu\footnote{For an electronic copy of the transparencies of this talk, please refer to the SUSY 2002 Web site: {\tt http://www.desy.de/$\tilde{~}$susy02/pl.7/landsberg.pdf.}}}
\date{}

\begin{document}

\maketitle

\begin{abstract}
As was suggested about a year ago, one of the most dramatic consequences of low-scale ($\sim 1$~TeV) quantum gravity is copious production of mini black holes at future accelerators and in ultra-high-energy cosmic ray collisions. Hawking radiation of these black holes is constrained mainly to our (3+1)-dimensional world and results in rich phenomenology. With the original idea having been cited over a hundred times since its appearance, we review the current status of astrophysical observations of black holes and selected topics in the mini black hole phenomenology, such as production rates at colliders and in cosmic rays, Hawking radiation as a sensitive probe of the dimensionality of extra space, as well as an exciting possibility of finding new physics in the decays of black holes.
\end{abstract}

\smallskip
{\flushright
{\it ``The Theory of Everything, if you dare to be bold,\\
Might be something more than a string orbifold.''}\\
Sheldon Glashow, 1986\rule{1.5in}{0pt}\\}

\section{Introduction}

The possibility that the universe has more than three spatial dimensions has been discussed since it was first suggested by Bernhard Riemann~\cite{Riemann}. What started as an abstract mathematical idea of a curved Riemannian space, soon became the foundation of the most profound physics theory of the last century, if not of the entire history of physics: Albert Einstein's general relativity~\cite{GR} (GR). While Einstein's theory was formulated in the three-plus-one space-time dimensions, it soon became apparent that the theory cannot be self-consistent up to the highest energies in its original form. In the 1920-ies, Theodor Kaluza and Oskar Klein~\cite{KK} showed that a unification of electromagnetism and general relativity is possible if the fifth, spatial dimension (compactified on a circle) is added to the four-dimensional space-time. While the original attempt has not led to a satisfactory common description of the two forces, with the rapid progress of string theory in the past quarter of century the concept of compact extra spatial dimensions regained its appeal. Based on the original idea by Kaluza and Klein, extra six or seven spatial dimensions in string theory are required for the most economical and symmetric formulation of its principles. In particular, string theory requires extra dimensions (ED) to establish its deep connection with the supersymmetry, which leads to the unification of gauge forces. String theory would have us believe that additional dimensions are compactified with the radii of the order of $10^{-32}$~m. 

In a new paradigm~\cite{add}, inspired by string theory (although not necessarily connected with it), Nima Arkani-Hamed, Savas Dimopoulos, and Georgi Dvali (ADD) suggested that several ($n$) of these compactified ED could be as large as $\sim 1$~mm. These {\it large extra dimensions\/} are introduced to solve the hierarchy problem of the standard model (SM) by lowering the Planck scale ($M_{\rm Pl}$) to a TeV energy range. (We further refer to this {\it fundamental\/} Planck scale as $M_P$.) In this new picture, the {\it apparent\/} Planck scale $M_{\rm Pl} = 1/\sqrt{G_N}$ only reflects the strength of gravity from the point of view of a three-dimensional observer. It is, in essence, just a virtual ``image'' of the fundamental, $(3+n)$-dimensional Planck scale, caused by an incorrect interpolation of the gravitational coupling (measured only at low energies) to high energies and short distances.

Since the original ADD paradigm had been proposed in 1998, numerous attempts to find large ED or constrain this model have been carried out. They include measurements of gravity at submillimeter distances~\cite{tabletop}, studies of various astrophysical and cosmological implications of large ED~\cite{astro}, and numerous collider searches for virtual and real graviton effects~\cite{collider}. For a detailed review of the existing constraints and sensitivity of future experiments, see, e.g. \cite{hs}. It is fair to say that the experimental measurements to date have nearly excluded only the case of two large ED\footnote{The case of a single extra dimension has been excluded from the very beginning, as it would require the size of this extra dimension to be of the order of the size of the solar system, in clear contradiction with the observations.}; for any larger number of them, the lower limit on the fundamental Planck scale is only $\sim 1$~TeV, hardly reaching the most natural range of scales expected in this model.

As was pointed out a few years ago~\cite{adm,bf,ehm}, an exciting consequence of TeV-scale quantum gravity is the possibility of production of black holes (BHs) at the accelerators. Recently, this phenomenon has been quantified for the case of TeV-scale particle collisions~\cite{dl,gt}, resulting in a mesmerizing prediction that future colliders would produce mini black holes at enormous rates (e.g. $\sim 1$~Hz at the LHC for $M_P = 1$~TeV), thus becoming black-hole factories. With the citation index of the original papers~\cite{dl,gt} now exceeding one hundred, the production of mini black holes in the lab became one of the most actively studied and rapidly evolving subjects in the phenomenology of models with extra dimensions over the past year.

In this talk we briefly review the situation with astronomical black hole observations and focus on phenomenology of the black hole production and decay in high-energy collisions. We point out exciting ways of studying quantum gravity and searching for new physics using large samples of black holes that may be accessible at future colliders and discuss the potential of the existing and future cosmic ray detectors for searches for black hole production in cosmic rays.

\section{Astronomical Black Holes}

While very few people doubt that black holes exist somewhere in the universe, perhaps even abundant, none of the astronomical black hole candidates found so far possess smoking-gun signatures uniquely identifying them as such. There are several ways astronomers look for black holes. For example, one of the first established black hole candidates, Cygnus X-1, was found by observing the orbital periods in superbright binary systems. The presence of a black hole as one of the two stars in Cygnus X-1 was inferred both from the large total power dissipated by the system and from extremely short time-scale of the intensity variations. Similar arguments lead astronomers to believe that quasars are powered by massive black holes. Furthermore, by observing X-ray flares in the active galactic nuclei (AGN), it is speculated that they are caused by large objects falling inside the AGN, being attracted by supermassive ($\sim 10^6$ solar masses) black holes located in their centers. There is even an evidence that such a supermassive black hole resides in the center of our own galaxy, the Milky Way. Clearly, these observations prove the presence of massive compact objects in many of the binary systems or AGN; however, proving that the critical density, necessary for a gravitational collapse into a black hole, has been achieved in these systems is rather complicated. The recent announcement of an observation of a ``strange'' star, apparently more dense than a neutron star~\cite{strangestar}, given the lack of a compelling cosmological explanation or even motivation for such an object, might indicate that there is a problem with the current methods of density estimates in extremely remote systems.

Other phenomena predicted for black holes in GR, such as frame dragging, have been observed as well. However, none of the existing black hole candidates have been tagged by several independent means so far. Given the large number of objects studied by the astronomers in their quest for black holes, it is questionable how unambiguous the single tags are.

Unfortunately, the most prominent feature of a black hole --- the Hawking radiation~\cite{Hawking}~--- has not been observed yet and is not likely to be ever observed by astronomical means. Indeed, even the smallest (and therefore the hottest) astronomical black holes with the mass close to the Chandrasekhar limit~\cite{Chandrasekhar}, have Hawking temperatures of only $\sim 100$~nK, which corresponds to the wavelength of Hawking radiation of $\sim 100$~km, and the total dissipating power of puny $\sim 10^{-28}$~W.\footnote{Note that the event horizon temperature of these black holes is much lower than the temperature of the CMB radiation, so at the present time the black holes are growing due to the accretion of relic radiation, rather than evaporating. They will start evaporating when the expanding universe cools down to temperatures below their Hawking temperature.}

Not only the event horizon of these black holes is colder than the lowest temperature ever achieved in the lab (the 1997 Nobel prize in physics was given to Steven Chu, Claude Cohen-Tannoudji, and William D. Phillips who reached the temperature as low as $\sim 1$~$\mu$K via optical cooling), but the wavelength of the radiation dissipated by a black hole is far from the visible spectrum and is resemblant of that of an AM radio station. Trying to detect such a radio broadcast with a vanishing transmitting power from thousands light-years away is but impossible. Given that the black hole dissipating power corresponds to only $\sim 100$ photons per second emitted by its entire event horizon and that the closest known black hole candidate is still over a thousand of light years away from us, not a single Hawking radiation photon ever hit our Earth since it has been formed! (In fact, one would have to wait $\sim 10^{14}$ years to observe a single photon from such a black hole to hit the Earth.) Thus, if the astronomical black holes were the only ones to exist, the Hawking radiation would be always just a theoretical concept, never testable experimentally.

While Hawking radiation would constitute a definite proof of the black hole nature of a compact object, there are other, indirect means of identifying the existence of the event horizon around it. It has been suggested~\cite{Narayan} that the lack of Type I X-ray bursts in the  binary systems, identified as black hole candidates, implies the presence of the event horizon. The argument is based on the fact that such X-ray bursts are frequent in the neutron star binaries, similar in size and magnitude to the black hole candidates. Consequently, if the black hole candidates did not have the event horizon, one would expect to see a similar X-ray burst activity, which contradicts the observations. An analogous argument applies to the X-ray supernovae that are believed to contain black holes. These supernovae are much dimmer than similar supernovae that are believed to contain neutron stars, which is considered to be an evidence for the presence of the event horizon in the black hole candidates. Unfortunately, an evidence based on the non-observation of a particular predicted phenomenon is inherently much more model-dependent and circumstantial than the one based on the observation of a certain effect. Perhaps, a more promising way to prove the existence of the event horizon in some of the black hole candidates is to compare the accretion disk shapes in various X-ray binaries~\cite{Narayan1}. Black hole binaries are expected to have advection-dominated accretion flows, drastically different from thin accretion disks, typical of subcritical binaries.

Probably, the best evidence for the existence of astronomical black holes would come from an observation of gravitational waves created in the collisions of two black holes, which LIGO and VIRGO detectors are looking for. However, current sensitivity of these interferometers is still short of the expected signal, even in the optimistic cosmological scenarios.

This leaves us with other places to look for black holes that are much smaller and consequently much hotter and easier to detect than their astronomical counterparts.

\section{Properties of Mini Black Holes}

Black holes are well understood general-relativistic
objects when their mass \mbh far exceeds the fundamental (higher
dimensional) Planck mass $\MP \sim 1$~TeV. As \mbh approaches \MP,
the BHs become ``stringy'' and their properties complex. In what 
follows, we will ignore this obstacle\footnote{Some of the properties of the stringy subplanckian ``precursors'' of black holes are discussed in Ref.~\protect\cite{de} and later in this review.}
and estimate the properties of light BHs by simple semiclassical 
arguments, strictly valid for $\MBH \gg \MP$. We expect that this 
will be an adequate approximation, since the important experimental
signatures rely on two simple qualitative properties: (i) the
absence of small couplings and (ii) the ``democratic" nature of BH 
decays, both of which may survive as average properties of the 
light descendants of BHs. We will focus on 
the production and sudden decay of Schwarzschild black holes.
The Schwarzschild radius $R_S$ of an $(4+n)$-dimensional black hole has been derived in Ref.~\cite{mp}, assuming that all $n$ extra dimensions are large ($\gg$ $R_S$).

As we expect unknown quantum gravity effects to play an increasingly 
important role for the BH mass approaching the fundamental Planck scale,
following the prescription of Ref.~\cite{dl}, we do not consider BH 
masses below the Planck scale. It is expected that the BH 
production rapidly turns on, once the relevant energy threshold
$\sim\! M_P$ is crossed. At lower energies, we expect BH production
to be exponentially suppressed due to the string excitations or 
other quantum effects. 

Note that the maximum center-of-mass energies accessible at the next generation of particle colliders and in ultrahigh-energy cosmic ray collisions are only a few TeV. Given the current lower constraints on the fundamental Planck scale of $\sim 1$~TeV, the artificial black holes that we might be able to study in the next decade will be barely transplanckian. Hence, the unknown quantum corrections to their GR properties are expected to be severe, and therefore we would like to focus on the most robust properties of these mini black holes that are expected to be affected the least by unknown effects of quantum gravity. Consequently, we do not consider spin and other black hole quantum numbers, as well as grey factors when discussing their production and decay, as their semiclassical form will be significantly modified by unknown quantum corrections. Later in this review we will discuss some of the subsequent attempts to take these BH properties into account.

\section{Black Hole Production and Decay}

Consider two partons with the center-of-mass energy $\sqrt{\hat s} =
\MBH$ colliding head-on. Semiclassical reasoning
suggests that if the impact parameter of the collision is less than the (higher dimensional) Schwarzschild radius, corresponding to this energy, a BH with the mass \mbh is formed. Therefore the total cross section of black hole production in particle collisions can be estimated from pure geometrical arguments and is of order $\pi R_S^2$.

Soon after Refs.~\cite{dl,gt} have appeared, Mikhail Voloshin suggested~\cite{Voloshin} an exponential suppression of the geometrical cross section based on the Gibbons-Hawking action~\cite{gh} argument. Detailed subsequent studies performed in simple string theory models~\cite{de}, using full GR calculations~\cite{GRcollisions}, or a path integral approach~\cite{path} did not confirm this finding and proved that the geometrical cross section is modified only by a numeric factor of order one. A flaw in the Gibbons-Hawking action argument of Ref.~\cite{Voloshin} was further found in Ref.~\cite{jt}: the use of this action implies that the black hole has been already formed, so describing the evolution of the two colliding particles before they cross the event horizon and form the black hole via Gibbons-Hawking action is not justified. By now there is a broad agreement that the production cross section is not significantly suppressed compared to a simple geometrical approximation, which we will consequently use through this review.

Using the expression for the Schwarzschild radius~\cite{mp}, we derive the following parton level BH production cross section~\cite{dl}:
$$
    \sigma(\MBH) \approx \pi R_S^2 = \frac{1}{M_P^2}
    \left[
      \frac{\MBH}{\MP} 
      \left( 
        \frac{8\Gamma\left(\frac{n+3}{2}\right)}{n+2}
      \right)
    \right]^\frac{2}{n+1}.
$$

In order to obtain the production cross section in $pp$ collisions at the LHC, 
we use the parton luminosity approach:\cite{dl,gt,EHLQ}
$$
    \frac{d\sigma(pp \to \mbox{BH} + X)}{d\MBH} = 
    \frac{dL}{dM_{\rm BH}} \hat{\sigma}(ab \to \mbox{BH})
    \left|_{\hat{s}=M^2_{\rm BH}}\right.,
$$
where the parton luminosity $dL/d\MBH$ is defined as the sum over
all the types of initial partons:
$$
    \frac{dL}{dM_{\rm BH}} = \frac{2\MBH}{s} 
    \sum_{a,b} \int_{M^2_{\rm BH}/s}^1  
    \frac{dx_a}{x_a} f_a(x_a) f_b(\frac{M^2_{\rm BH}}{s x_a}),
$$
and $f_i(x_i)$ are the parton distribution functions (PDFs). We
used the MRSD$-'$~\cite{MRSD} PDFs with the $Q^2$ scale taken
to be equal to \MBH, which is within the allowed range of these
PDFs for up to the kinematic limit at the LHC. The dependence of 
the cross section on the choice of PDF is $\sim 10\%$.
The total production cross section for $\MBH > M_P$ at the LHC, 
obtained from the above equation, ranges between 15 nb and 1 pb for 
the Planck scale between 1 TeV and 5 TeV, and varies by $\approx 10\%$ 
for $n$ between 2 and 7.

Once produced, mini black holes quickly evaporate via Hawking 
radiation~\cite{Hawking} with a characteristic temperature
$$
    T_H = \MP
    \left(
      \frac{\MP}{\MBH}\frac{n+2}{8\Gamma\left(\frac{n+3}{2}\right)}
    \right)^\frac{1}{n+1}\frac{n+1}{4\sqrt{\pi}} = \frac{n+1}{4\pi R_S}
$$
of $\sim 100$~GeV~\cite{dl,gt}. The average multiplicity of particles 
produced in the process of BH evaporation is given by~\cite{dl,gt} and 
is of the order of half-a-dozen for typical BH masses accessible 
at the LHC. Since gravitational coupling is flavor-blind, a BH emits all the 
$\approx 120$ SM particle and antiparticle degrees of freedom with 
roughly equal probability. Accounting for color and spin, we expect 
$\approx 75\%$ of particles produced in BH decays to be quarks and gluons, 
$\approx 10\%$ charged leptons, $\approx 5\%$ neutrinos, and $\approx 5\%$ 
photons or $W/Z$ bosons, each carrying hundreds of GeV of energy. 
Similarly, if there exist new particles with the scale $\sim 100$~GeV, 
they would be produced in the decays of BHs with the probability 
similar to that for SM species. For example, a sufficiently light 
Higgs boson is expected to be emitted in BH decays with $\sim 1\%$ probability. This has exciting consequences for searches for new physics at the LHC and beyond, as the production cross section for any new particle via
this mechanism is (i) large, and (ii) depends only weakly on particle mass, 
in contrast with the exponentially suppressed direct production mechanism.

A relatively large fraction of prompt and energetic photons, electrons, 
and muons expected in the high-multiplicity BH decays would 
make it possible to select pure samples of BH events, which are also 
easy to trigger on~\cite{dl,gt}. At the same time, only a small fraction 
of particles produced in the BH decays are undetectable gravitons 
and neutrinos, so most of the BH mass is radiated in the form of visible 
energy, making it easy to detect.

It has been recently argued~\cite{chromosphere} that the fragmentation of quarks and jets emitted in the black hole evaporation might be significantly altered by the presence of a dense and hot QCD plasma (``chromosphere'') around the event horizon. If this argument is correct, one would expect much softer hadronic component in the black hole events. However, we would like to point out that one would still have a significant number of energetic jets due to the decay of weakly interacting $W/Z$ and Higgs bosons, as well as tau leptons, emitted in the process of BH evaporation and penetrating the chromosphere before decaying into jetty final states. In any case, tagging of the black hole events by the presence of an energetic lepton or a photon and large total energy deposited in the detector is a fairly model-independent approach.

\begin{figure}[tbp]
\begin{center}
\epsfxsize=5in
\epsffile{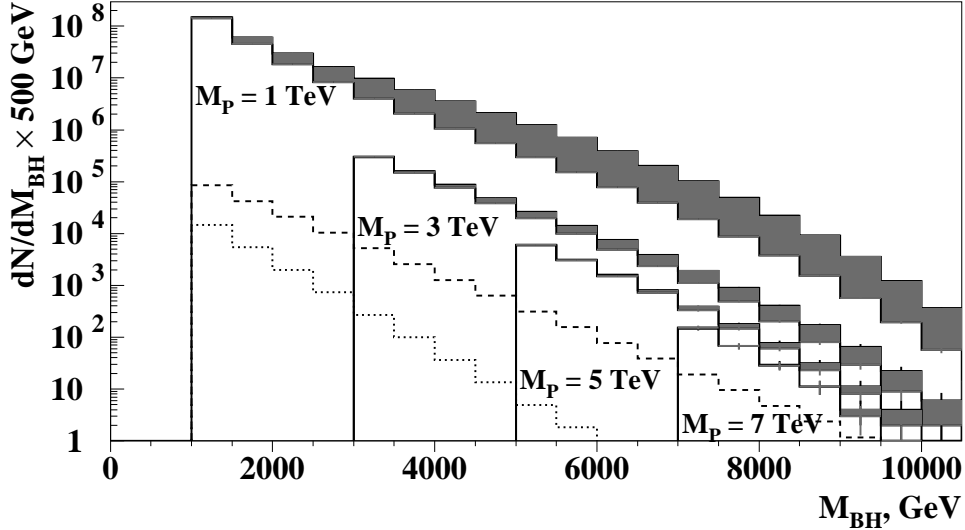}
\caption{Number of BHs produced at the LHC in the electron or photon decay 
channels, with 100~\protect\ifb of integrated luminosity, as a function of the BH 
mass. The shaded regions correspond to the variation in the number of events 
for $n$ between 2 and 7. The dashed line shows total SM background 
(from inclusive $Z(ee)$ and direct photon production). The dotted line 
corresponds  to the $Z(ee)+X$ background alone. From Ref.~\protect\cite{dl}}
\label{nbh}
\end{center}
\end{figure}

In Fig.~\ref{nbh} we show the number of BH events tagged by the presence of an energetic electron or photon among the decay products in, 100~fb$^{-1}$ of data collected at the LHC, along with SM backgrounds, as a function of the BH mass~\cite{dl}. It is clear that 
very clean and large samples of BHs can be produced at the LHC up to Planck scale of $\sim 5$ TeV. Note that the BH discovery potential at the LHC is maximized in the
$e/\mu+X$ channels, where background is much smaller than that 
in the $\gamma+X$ channel (see Fig.~\ref{nbh}). The reach of a simple 
counting experiment extends up to $\MP \approx 9$ TeV ($n=2$--7),
for which one would expect to see a handful of BH events with negligible 
background. 

\begin{figure}[htbp]
\begin{center}
\epsfxsize=4in
\epsffile{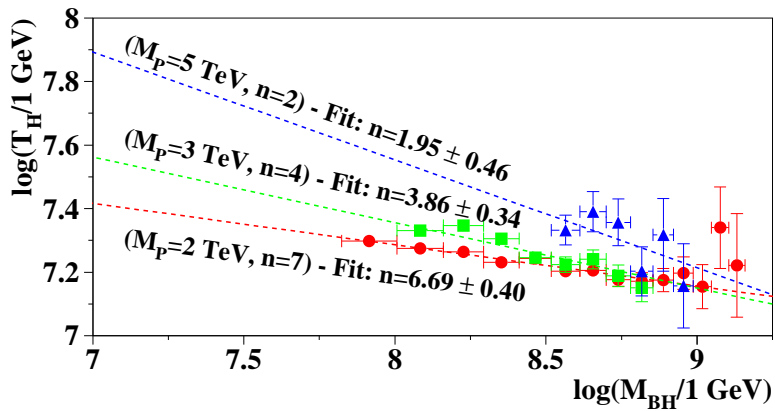}
\caption{Determination of the
dimensionality of  space via Wien's displacement law at the LHC with
100~\protect\ifb\ of data. From Ref.~\protect\cite{dl}.}
\label{Wien}
\end{center}
\end{figure}

A sensitive test of properties of Hawking radiation can be performed by measuring the relationship between the mass of the BH (reconstructed from the total energy of all the decay products) and its Hawking temperature (measured from the energy spectrum of the electron or photon tags). One can use the measured \mbh vs. \TH\ dependence to determine both the fundamental Planck scale \mp and the dimensionality of space $n$. This is a multidimensional equivalent of the Wien's law. Note that the dimensionality of extra space can be determined in a largely model-independent way via taking a logarithm of both parts of the expression for Hawking temperature: $\log(T_H) = -\,\frac{1}{n+1}\log(\MBH) + \mbox{const}$, 
where the constant does not depend on the BH mass, but only
on \mp and on detailed properties of the bulk space, such as the shape of
extra dimensions~\cite{dl}. Therefore, the slope of a straight-line fit to
the $\log(T_H)$ vs. $\log(\MBH)$ data offers a direct way of determining
the dimensionality of space. The reach of this method at the LHC is illustrated in Fig.~\ref{Wien} and discussed in detail in Ref.~\cite{dl}. Note that the determination of the dimensionality of space-time by this method is fundamentally different from other ways of determining $n$, e.g.
by studying a monojet signature or a virtual graviton exchange processes,
also predicted by theories with large extra dimensions. The latter always depend on the volume of extra space, and therefore cannot provide with the direct way of measuring $n$ without making assumptions about the relative size of large extra dimensions. The former, on the other hand, depends only on the area of the event horizon of a black hole, which does not depend on the size of large extra dimensions or their shape.

An interesting possibility studied in Ref.~\cite{de} is production of a precursor of a black hole, i.e. a long and jagged highly-exited string, dubbed as a ``string ball'' due to its folding via a random walk. As shown in Ref.~\cite{de}, there are three characteristic string ball production regimes, depending on the mass of the produced excitation $M$, the string scale $M_S < M_P$,  and the string coupling $g_s < 1$. For $M_S < M < M_S/g_s$, the production cross section increases $\propto M^2$, until it reaches saturation at $M \sim M_S/g_s$ and stays the same up to the string ball mass $\sim M_S/g_s^2$, when the black hole is formed and the cross section agrees with that from Ref.~\cite{dl}.
A string ball has properties similar to those of a black hole, except that its evaporation temperature, known as Hagedorn temperature~\cite{Hagedorn}, is constant: $T_S = M_S/(2\sqrt{2}\pi)$. Thus, the correlation between the temperature of the characteristic spectrum and the string ball mass may reveal the transition from the Hagedorn to Hawking regime, which can be used to estimate $M_S$ and $g_s$. Another possibility is a production of higher-dimensional objects, e.g. black $p$-branes, rather than spherically symmetric black holes ($p=0$)~\cite{blackbranes}. For a detailed review see, e.g. Ref.~\cite{kingman}.

Several of the recent papers looked at the Hawking evaporation process in more detail. While we do not believe that significant refinement of the semiclassical approximation used in~\cite{dl} is possible without knowing the underlying quantum theory of gravity, we list some of these attempts. The effects of using the microcanonical ensemble approach, which takes into account that the energy of the emitted particles is comparable to the black hole mass, have been discussed in Ref.~\cite{microcanonical} and generally result in the increased black hole lifetime. The recoil effect in the evaporation has been studied~\cite{recoil} as well. A number of attempts to calculate grey-body factors and to take into account the black hole spin have been made; however all of them hinge on a simple semiclassical approximation, often only in four space-time dimensions. Consequently, these results are  unlikely to survive quantum gravity corrections, so we choose not to discuss them here in more detail.

\section{Discovering New Physics in the Decays of Black Holes}

As was mentioned earlier, new particles with the mass $\sim 100$~GeV would be produced in the process of black hole evaporation with a relatively large probability: $\sim 1\%$ times the number of their quantum degrees of freedom. Consequently, it may be advantageous to look for new particles among the decay products of black holes in large samples accessible at the LHC and other future colliders.

As an example~\cite{gl}, we study the discovery potential of the BH sample collected at the LHC for a SM-like Higgs boson with the mass of 130 GeV, still allowed in low-scale 
supersymmetry models, but very hard to establish at the Fermilab Tevatron~\cite{SUSYHiggs}. We consider the decay of the Higgs boson into a pair of jets (with the branching fraction of 67\%), dominated by the $b\bar b$ final state (57\%), with an additional 10\% contribution from the $c\bar c$, $gg$, and hadronic $\tau\tau$ final states.

We model the production and decay of the BH with the TRUENOIR Monte Carlo 
generator~\cite{Snowmass}, which implements a heuristic algorithm to describe a spontaneous decay of a BH. The generator is interfaced with the PYTHIA Monte 
Carlo program~\cite{PYTHIA} to account for the effects of initial and final state radiation, particle decay, and fragmentation.\footnote{The black hole production and decay is also being implemented in HERWIG~\protect\cite{HERWIG}.} We used a 1\% probability 
to emit the Higgs particle in the BH decay. We reconstruct final state particles within the acceptance of a typical LHC detector and smear their energies with the expected resolutions. 

\begin{figure}[tbp]
\begin{center}
\epsfxsize=3.8in
\epsffile{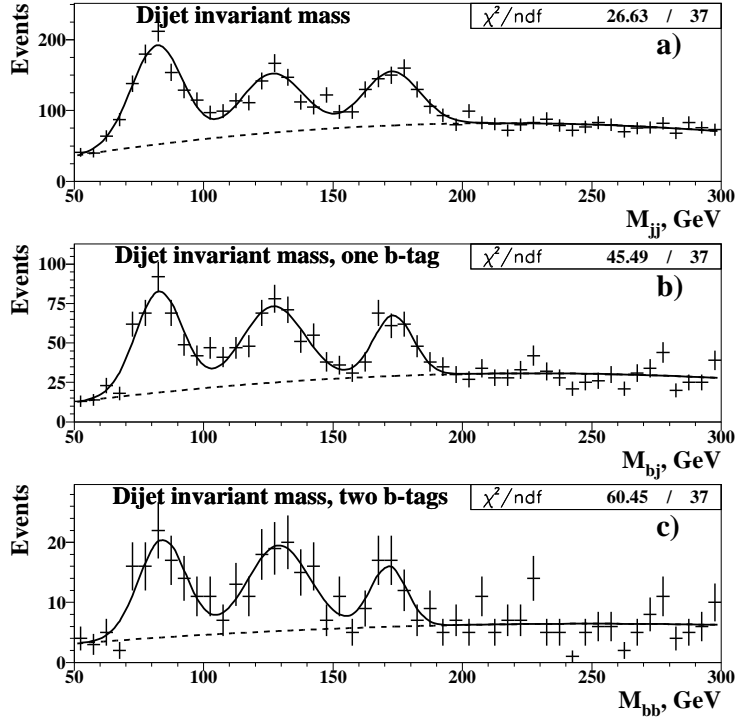}
\caption{Dijet invariant mass observed in the BH decays with a prompt 
lepton or photon tag in $\approx$3~pb$^{-1}$ of the LHC data, for 
$M_P = 1$~TeV and $n=3$: (a) all jet combinations; (b) jet combinations 
with at least one of the jets tagged as a $b$-jet; (c) jet 
combinations with both jets tagged as $b$-jets. The solid line is a 
fit to a sum of three Gaussians and a polynomial background (shown 
with the dashed line). The three peaks correspond to the $W/Z$ bosons, 
the Higgs boson, and the top quark (see text). The $\chi^2$ per d.o.f.
is shown to demonstrate the quality of the fit. Note that as the 
maximum likelihood fit was used for all cases, the $\chi^2$ in (c) is 
not an appropriate measure of the fit quality due to low statistics. 
Using the Poisson statistics, the probability of the fit (c) is 8\%. 
From Ref.~\protect\cite{gl}.}
\label{bhhiggs}
\end{center}
\end{figure}

The simplest way to look for the Higgs boson in the BH decays is to use the 
dijet invariant mass spectrum for all possible combinations of jets found among the final state products. This spectrum is shown in Fig.~\ref{bhhiggs} for $M_P = 1$~TeV and $n=3$. The three panes correspond 
to all jet combinations (with the average of approximately four jet 
combinations per event), combinations with at least one $b$-tagged jet, 
and combinations with both jets $b$-tagged. (We used typical tagging efficiency and mistag probabilities of the LHC detector to simulate $b$-tags.)

The most prominent feature 
in all three plots is the presence of three peaks with the masses around 
80, 130, and 175 GeV. The first peak is due to the hadronic decays of the 
$W$ and $Z$ bosons produced in the BH decay either directly or via the  
the top and Higgs decays. (The resolution of a typical LHC detector 
does not allow to resolve the $W$ and $Z$ in the dijet decay mode.) 
The second peak is due to the $h \to jj$ decays, and the third peak is due to the $t \to  Wb \to jjb$ decays, where the top quark is highly boosted. In this case, one of the jets from the $W$ decay sometimes overlaps with the prompt 
$b$-jet from the top quark decay, and thus the two are reconstructed as 
a single jet; when combined with the second jet from the $W$ decay, this 
gives a dijet invariant mass peak at the top quark mass. The data set
shown in Fig.~\ref{bhhiggs} corresponds to 50K BH events, which, given the 15~nb production cross section for $M_P = 1$~TeV and $n = 3$, is equivalent to the integrated luminosity of 3 pb$^{-1}$, or less than an hour of the LHC operation at the nominal instantaneous luminosity. The significance of the Higgs signal shown in Fig.~\ref{bhhiggs}a is 6.7$\sigma$, even without $b$-tagging involved.

With this method, a $5\sigma$ discovery of the 130 GeV Higgs boson may be possible with ${\cal L} \approx 2$~pb$^{-1}$ (first day), 100~pb$^{-1}$ (first week), 1~fb$^{-1}$ (first month), 10~fb$^{-1}$ (first year), and 100~fb$^{-1}$ (one year at the nominal luminosity) for the fundamental Planck scale of 1, 2, 3, 4, and 5~TeV, respectively, even with incomplete and poorly calibrated  detector. The integrated luminosity required is significantly lower than that for discovery in the direct production, if the Planck scale is below 4 TeV. 

While this study was done for a particular value of the Higgs boson mass, the dependence of the new approach on the Higgs mass is small. Moreover, this approach is applicable to searches for other new particles with the masses $\sim 100$~GeV, e.g. low-scale supersymmetry. Light slepton or top squark searches via this technique may be particularly fruitful. Very similar conclusions apply not only to BHs, but to intermediate quantum states, such as string balls~\cite{de}, which have similar production cross section 
and decay modes as BHs. In this case, the relevant mass scale is not the 
Planck scale, but the string scale, which determines the Hagedorn evaporation temperature.

Large sample of black holes accessible at the LHC can be used even to study some of the properties of known particles, see, e.g. Ref.~\cite{uehara}.

\section{Black Holes in Cosmic Rays}

\begin{figure}[thbp]
\begin{center}
\epsfxsize=3.5in
\epsffile{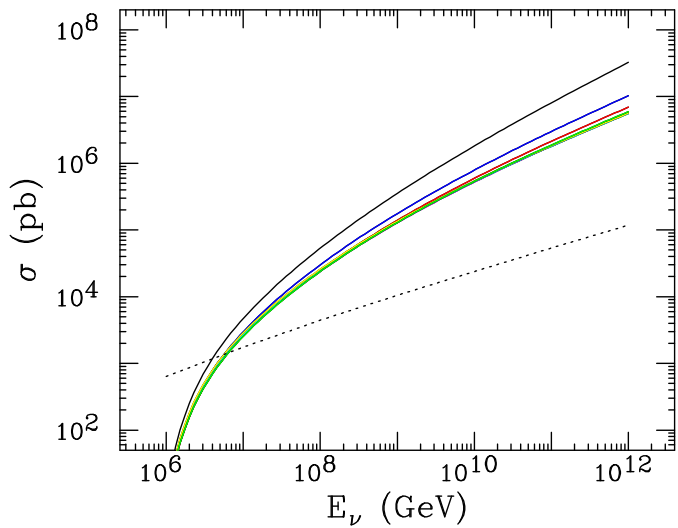}
\caption{Cross sections $\sigma ( \nu N \to {\rm BH})$ for $M_P =
M_{\rm BH}^{\rm min} = 1$~TeV and $n=1, \ldots, 7$. (The last four
curves are virtually indistinguishable.)  The dotted curve is for the
SM process $\nu N \to \ell X$. From Ref.~\protect\cite{Feng}.}
\label{bhcr}
\end{center}
\end{figure}

Recently, it has been suggested that mini black hole production can be also observed in the interactions of ultra-high-energy neutrinos with the Earth or its atmosphere~\cite{Feng}. For neutrino energies above $\sim 10^7$ GeV, the BH production cross section in $\nu q$ collisions would exceed their SM interaction rate (see Fig.~\ref{bhcr}). Several ways of detecting BH production in neutrino interactions have been proposed~\cite{Feng,CR,afgs,IceCube}, including large-scale ground-based arrays, space probes, and neutrino telescopes as detecting media. 

If the fundamental Planck scale is sufficiently low (1--3 TeV), up to a hundred of BH events could be observed by, e.g. Pierre Auger observatory even before the LHC turns on. These estimates are based on the so-called guaranteed, or cosmogenic neutrino flux~\cite{flux}. In certain cosmological models, this flux could be significantly enhanced by additional sources of neutrino emission, e.g. AGN; in this case even larger event count is possible.

There are two ways to tell the neutrino interaction that results in a black hole formation from the standard model processes. The first is based on a particular particle content in the black hole events, and would require good particle identification, perhaps beyond the capabilities of the existing detectors. The second approach is based on the comparison of the event rate for Earth skimming neutrinos (i.e., those that traverse the Earth crust via a short chord, close to the surface) with that for the quasi-horizontal neutrinos (i.e., those that do not penetrate the Earth, but traverse the atmosphere at a small angle). In the former case, many of the neutrinos would be stopped in the Earth due to the large cross section of BH production. That would suppress the rate of the Earth-skimming-neutrino events in a typical ground array detector, such Pierre Auger. At the same time, the rate of the quasi-horizontal events would increase, as the total cross section, which governs this rate, is dominated by the black hole production and therefore is higher than in the SM case. By measuring the ratio of the two rates, it is possible to distinguish the standard model events from the black hole production even with a handful of detected events~\cite{afgs}.

Another interesting observation can be done with the IceCube large-scale neutrino telescope at the South Pole, by looking at the zenith angle dependence of the neutrino events at various incidental energies. Similar to the previous argument, significant reduction of the number of observed events due to the neutrino absorption via black hole production in the Earth material surrounding the detector, would occur at smaller zenith angles than that in the case of the SM-only neutrino interactions~\cite{IceCube}. In addition, particle identification capabilities of the IceCube detector are likely to make it possible to detect the black hole events directly by looking at the event shape. While it has not been mentioned in the original papers~\cite{IceCube}, we would like to note an additional azimuthal dependence of the event rate for high-energy Earth-skimming neutrinos in the IceCube due to the presence of several mountain ridges near the South Pole (particularly, the Transantarctic Mountain Ridge). These mountains are not thick enough to significantly reduce the flux of Earth skimming high-energy neutrinos due to the standard model interactions, but are sufficiently thick to absorb these neutrinos if the black hole creation is allowed. That would constitute a spectacular signature.

\section{Reentering Black Holes}

A new interesting topic in black hole phenomenology is the possibility that a black hole, once produced, moves away into the bulk space. Normally it does not happen as the black holes produced in collisions at the LHC or in cosmic ray interactions are likely to have charge, color, or lepton/baryon number hair that would keep them on the brane. However, an exciting possibility of that kind is allowed in the case when the strength of gravity in the bulk and on the brane is very different. This is the case, e.g. in the ADD scenario with an additional brane term~\cite{braneterm}, or in the case of infinite-volume extra dimensions~\cite{infinite}. 

In these models, a particle produced in a subplanckian collision, e.g. a graviton, could move away in the bulk, where it becomes a black hole due to much lower effective Planck scale in extra dimensions. Since the Planck scale in the bulk is very low, e.g. $\sim 0.01$~eV in the infinite-volume scenario~\cite{infinite}, the newly-formed black hole is very cold and therefore essentially stable. Furthermore, it generally does not move far away from the brane due to gravitational attraction to it, and can further accrete mass from relic energy density in the bulk and from other particles produced in the subsequent collisions. Once the mass of the black hole reaches the mass of the order of the apparent Planck scale, $M_{\rm Pl} \sim 10^{19}$~GeV, the event horizon of the bulk black hole grows so large that it touches the brane, and the black hole immediately evaporates on the brane into $\sim 10$ particles with the energy $\sim 10^{18}$ GeV each. (The energy released in such an event is similar to that in an explosion of a large, few hundred pound conventional bomb!) If such black holes are copiously produced by a remote cosmic accelerator of a reasonable energy, they could act as a source of the highest-energy cosmic rays that are emitted in the process of decay and deceleration of the super-energetic BH remnants.

Even if the mass of the black hole in the bulk is small, it has certain probability to reenter our brane. In this case, since the event horizon cannot be destroyed, once it has been formed, such a subplanckian object would likely to act as a black hole on the brane and evaporate similarly to a transplanckian black hole discussed above.

The details of these exciting preliminary observations are under investigation and will be reported shortly~\cite{dgl}.

\section{Conclusions}

To conclude, black hole production at the LHC and in cosmic rays may be one of the early
signatures of TeV-scale quantum gravity. It has three advantages:
\begin{enumerate}
\item Large Cross Section. Because no small dimensionless coupling constants, analogous to $\alpha$, suppress the production of black holes. This leads to enormous rates.

\item Hard, Prompt, Charged Leptons and Photons. Because thermal decays are flavor-blind. This signature has practically vanishing SM background.

\item Little Missing Energy. This facilitates the determination of the mass and the temperature of the black hole, and may lead to a test of Hawking radiation.
\end{enumerate}

Large samples of black holes accessible by the LHC and the next generation of colliders would allow for precision determination of the parameters of the bulk space and may even result in the discovery of new particles in the black hole evaporation. Limited samples of black hole events may be observed in ultra-high-energy cosmic ray experiments, even before the LHC turns on.

If large extra dimensions are realized in nature, the production and detailed studies of black holes in the lab are just few years away. That would mark an exciting transition for astroparticle physics: its true unification with cosmology~--- the ``Grand Unification'' to live for.

\section*{Acknowledgments}

I would like to thanks the SUSY 2002 organizers for a kind invitation, warm hospitality, and for an exciting and inspiring conference. I am indebted to my coauthor on the first black hole paper~\cite{dl}, Savas Dimopoulos, for a number of stimulating discussions and support. The preliminary results on the reentering black holes are based on numerous communications with Gia Dvali. I am grateful to Jonathan Feng for several clarifications on the cosmic ray detection of black holes and to Robert Brandenberger for comments on the astrophysical black hole section. This work has been partially supported by the U.S.~Department of Energy under Grant No. DE-FG02-91ER40688 and by the Alfred P. Sloan Foundation.

\end{document}